\documentclass[aps,prc,twocolumn,showpacs,superscriptaddress,nofootinbib,nolongbibliography,floatfix]{revtex4-2}  
\usepackage{changes} 
\usepackage[para,online,flushleft]{threeparttable}
\usepackage{graphicx}  
\usepackage{dcolumn}   
\usepackage{bm}        
\usepackage{amssymb}   
\usepackage{amssymb,amsmath,amsfonts} 
\usepackage[breaklinks=true,debug=true]{hyperref}
\usepackage{braket}    
\usepackage{booktabs}
\usepackage{multirow}
\usepackage{adjustbox}
\usepackage{gnuplottex}
\usepackage[T1]{fontenc}
\usepackage[utf8]{inputenc}
\usepackage{color}
\usepackage{dsfont}
\usepackage{tikz}
\usepackage{pgfplots}
\usepgfplotslibrary{fillbetween}
\usetikzlibrary {arrows.meta}

\usepgfplotslibrary{groupplots}
\pgfplotsset{compat=newest}
\usepgfplotslibrary{units}
\usetikzlibrary{patterns}

\begin{document}


\title{New shell-model calculations of the $\delta_C$ correction to superallowed $0^+\rightarrow0^+$ nuclear $\beta$ decay and standard-model implications}
\author{L. Xayavong} 
\affiliation{Department of Physics, Yonsei University, Seoul 03722, South Korea}
\affiliation{LP2IB (CNRS/IN2P3-Universit\'e de Bordeaux), 33170 Gradignan cedex, France}
\author{N. A. Smirnova}
\affiliation{LP2IB (CNRS/IN2P3-Universit\'e de Bordeaux), 33170 Gradignan cedex, France}
\author{F. Nowacki}
\affiliation{IPHC, IN2P3-CNRS/Universit\'e Louis Pasteur BP 28, F-67037 Strasbourg Cedex 2, France}
\date{\today}

\begin{abstract} 
Refined calculations of the radial mismatch correction, $\delta_{C2}$, to superallowed $0^+\rightarrow0^+$ nuclear $\beta$ decay are performed using the shell model with realistic Woods-Saxon radial wave functions. Two important improvements are introduced: i) charge radii used to constrain the length parameter are evaluated within a generalized formula, where proton occupation numbers are substituted by sums of spectroscopic factors, while radial wave functions are required to match separation energies with respect to the intermediate $(A-1)$-nucleon states by adjusting parameters such as the potential depth; ii) configuration mixing wave functions and energies for many-particle states are obtained through the diagonalization of well-established effective interactions in large configuration spaces without truncation. Furthermore, a variation of $\pm0.1$\,fm in the surface diffuseness parameter is now incorporated as a source of uncertainty. 
The present results are generally in fairly good agreement with those from previous studies. As an exception, the $\delta_{C2}$ value obtained for $^{18}$Ne is smaller by approximately a factor of two, principally due to the updated charge-radius treatment. A reduction is also observed in most cases with $A\ge38$, through the deviations generally remain within the newly assigned error bars. The smaller isospin-mixing counterpart, $\delta_{C1}$, is strongly interaction-dependent, roughly following an inverse-square law with respect to the energy separation between the lowest admixed levels. Therefore, an additional procedure to ensure isobaric displacements within the isospin multiplets appears to be indispensable. 
Our results for $\delta_{C2}$ lead to a new averaged $\overline{\mathcal{F}t}$ value of $3073.11(99)_{stat}(36)_{\delta_R'}(173)_{\delta_{NS}}$~s with $\chi^2/\nu=0.624$. The corresponding $|V_{ud}|$ value is 0.97359(33). 
\end{abstract} 

\maketitle

\section{Introduction}\label{intro}

Superallowed $0^+\rightarrow0^+$ nuclear $\beta$ decay plays an important role in probing the fundamental symmetries underlying the Standard Model. As a pure Fermi process, this transition, under isospin SU(2) symmetry, connects only isobaric analog states, and its nuclear matrix element is model-independent. 
Despite this simplicity, several theoretical corrections are required in order to compare the predictions from the Standard Model with the high-precision results from experiments. The master formula for this process is expressed as, 
\begin{equation}\label{eq1}
\displaystyle \mathcal{F}t = ft(1+\delta_{R}')(1+\delta_{NS}-\delta_C) = \frac{K}{2G_V^2(1+\Delta_R^V)},
\end{equation}
where $K$ is a known constant~\cite{HaTo2020}, $G_V$ is the vector coupling constant, and $ft$ is the product of the statistical rate function ($f$) and the partial half-life ($t$). Furthermore, $\delta_C$ is the correction due to isospin-symmetry breaking, which is the core interest of the present study. Additionally, the three remaining theoretical correction terms account for radiative effects~\cite{HaTo2020}, with
$\Delta_R^V$ being nucleus-independent~\cite{PhysRevD.100.073008,PhysRevLett.132.191901,Chien2018}, 
$\delta_R'$ depending only on the atomic number of daughter nucleus and the decay $Q$ value\,\cite{HaTo2020}, and $\delta_{NS}$, which is nuclear structure-dependent\,\cite{PhysRevLett.133.211801,PhysRevLett.134.012501,PhysRevC.107.035503,ToHa1977}. 

According to the conserved vector current (CVC) hypothesis, the corrected $\mathcal{F}t$ must be independent of the nuclear medium. Importantly, fluctuations in $\mathcal{F}t$ can be attributed to the presence of weak exotic currents not accounted for in the Standard Model\,\cite{HaTo2020}. This property can be verified using experimental $ft$ values and calculated values of the theoretical corrections. Additionally, with CVC validated, the value for the vector coupling constant, $G_V$, deduced from Eq.\,\eqref{eq1}, when combined with the value obtained from the purely leptonic $\mu$ decay\,\cite{PhysRevD.87.052003}, yield the Kobayashi-Maskawa mixing angle between $u$ and $d$ quarks, $|V_{ud}|$. This matrix element is an essential ingredient for testing the top-row unitarity of the Cabibbo–Kobayashi–Maskawa (CKM) matrix~\cite{HaTo2020}. Currently, the relative precision for $ft$ has reached $10^{-4}$~\cite{HaTo2020} in 15 cases, covering the mass range from $^{10}$C to $^{74}$Rb. The primary limiting factors in these tests are the theoretical corrections, particularly $\delta_C$ and $\delta_{NS}$, although their magnitudes are typically below 1\,\%. 

\begin{table*}[ht!]
\centering
\caption{Comparison of model spaces and effective interactions used in this work and in the calculations by Towner and Hardy (TH)\,\cite{ToHa2002,ToHa2008,HaTo2020}. 
A severe truncation was imposed in the TH calculations, except in the $p$ and $sd$ shells. 
}
\begin{threeparttable}
\begin{ruledtabular}
\begin{tabular}{c|c|c|c|c|c|c}
\multirow{2}{*}{Emitters}  &  \multicolumn{2}{c|}{Model space} &  \multicolumn{2}{c|}{$H_0$}   & \multicolumn{2}{c}{$V_{INC}$} \\
\cline{2-7}
  &  This work & TH  & This work & TH & This work & TH\tnote{d} \\
\hline
$^{10}$C & $p$ & $p$ &  (CKPOT, CKI, CKII)\,\cite{Cohen1967,Cohen1965} & CKPOT, PWBT\,\cite{PhysRevC.46.923}  & OB~\cite{OrBr1989x} &  \\
$^{14}$O & $p$ & $psd$ &  CKPOT, CKI, CKII & $p+sd+psd$\tnote{b}  & OB~\cite{OrBr1989x} &  \\
$^{18}$Ne, $^{22}$Mg & ZBM & $psd$ &  REWIL\,\cite{PhysRevC.7.974}, (ZBM, ZBMI, ZBMII)\,\cite{PhysRevLett.21.39} & $p+sd+psd$\tnote{b} & OB~\cite{OrBr1989x} &  \\
$^{26}$Al, $^{26}$Si & $sd$ & $psd$ & USD\,\cite{USD}, (USDA, USDB)\,\cite{USDab} & $p+sd+psd$\tnote{b} & OB~\cite{OrBr1989x} &  \\
$^{30}$S, $^{34}$Cl, $^{34}$Ar & $sd$ & $sd$ & USD, USDA, USDB & USD, USDA, USDB & OB~\cite{OrBr1989x} &  \\
$^{38}$K to $^{46}$V & ZBM2 & $sdpf$ & ZBM2m\,\cite{Bis2014} & $sd+pf+sdpf$\tnote{c} & IVSPE+Coul &  \\
$^{50}$Mn to $^{62}$Ga &  $pf$  & $pf$  & GXPF1A\,\cite{gx1a}, KB3G\,\cite{kb3g}, FPD6\,\cite{fpd6} & KB3G, FPD6\tnote{a}, GXPF1A & OB~\cite{OrBr1989x} &  \\
$^{66}$As, $^{70}$Br  & $jj44$ & $pf$ & JUN45\,\cite{jun45}, MRG\,\cite{mrg} & KB3G, FPD6\tnote{a}, GXPF1A & IVSPE+Coul &  \\
\end{tabular}
\end{ruledtabular}
\begin{tablenotes}
{ \raggedright
\item[a] A modification to the monopole component was introduced. \\
\item[b] Combination of $p$- and $sd$-orbital two-body matrix elements (TBMEs), while $psd$ cross-shell elements are taken from Ref.\,\cite{MILLENER1975315}. \\
\item[c] Combination of $sd$- and $pf$-orbital TBMEs, while the $sdpf$ cross-shell elements are taken from Ref.\,\cite{MILLENER1975315}. \\
\item[d] Details are given in the main text. \\
}
\end{tablenotes}
\end{threeparttable}
\label{tab1}
\end{table*}

Over the past half-century, extensive efforts have been devoted to improving $\delta_C$ calculations. Among the diverse sets of theoretical approaches~\cite{XaNa2018,PhysRevC.105.044308,PhysRevC.109.014317,ToHa2008,OrBr1985,Li2009,Sat2011,Au2009,Dam1969}, only the shell model with realistic Woods-Saxon (WS) radial wave functions~\cite{ToHa2002,ToHa2008,HaTo2015,HaTo2020,XaNa2018} provides consistent results, supporting the standard electroweak theory. Certainly, this consistency alone would not be sufficient to validate the calculated $\delta_C$ values, since the Standard Model itself is still under scrutiny. Nevertheless, the degree of disagreement observed within the other theoretical approaches generally appears too severe to justify any claims of new physics. In addition to costly experimental tests of $\delta_C$, such as those discussed in Refs.\,\cite{XaNa2018,HaTo2020,PhysRevLett.107.182301,PhysRevLett.73.396,PhysRevC.102.054325,PhysRevC.87.064306,PhysRevLett.112.102502}, it is equally important to investigate potential shortcomings in the applied theoretical models. 
In this article, we present new shell-model calculations of $\delta_C$, with a special focus on its radial mismatch component, using an updated optimization procedure for WS parameters, where separation energies and charge radii are consistently fitted\,\cite{XaNa2018,Xthesis}. 
For all transitions, we consider the largest possible configuration spaces without truncation through large-scale shell-model calculations. We also incorporate recent developments in effective interactions, in particular the modified ZBM2~\cite{Bis2014}, which offers a suitable description of nuclei in the vicinity of $^{40}$Ca, as well as the JUN45~\cite{jun45} and MRG~\cite{mrg}, which have been very successful in the upper $pf$-shell region. 

\section{Formalism}\label{form}

Within the phenomenological shell model\,\cite{OrBr1985,OrBr1989x,physics5020026}, the effective Hamiltonian is written as 
\begin{equation}
    H=H_0 + V_{INC}, 
\end{equation}
where the first term conserves isospin, $[H_0,\bm{T}]=0$, while the second term does not, $[V_{INC},\bm{T}]\ne0$. Briefly speaking, the part $H_0$ is typically obtained through a least-squares fit of two-body matrix elements to available binding and excitation energies. In addition, the isoscalar single-particle energies are replaced with isospin-averaged experimental values, extracted from corresponding one-particle and one-hole states in the vicinity of closed-shell nuclei. 
The isospin nonconserving counterpart $V_{INC}$ is separately determined. 
According to the seminal works by Ormand and Brown (OB)\,\cite{OrBr1989x}, this component is modeled as the sum of the two-body Coulomb interaction and Yukawa-type potentials representing charge-dependent nuclear forces. 
Apart from the one-body contribution which can be directly substituted by isovector single-particle energies (IVSPE), the parameters for these interactions are constrained by experimental data on isotopic mass shifts, which are primarily sensitive to the $b$ and $c$ coefficients of the isobaric multiplet mass equation (IMME). 

The presence of $V_{INC}$ gives rise to an isospin impurity in many-particle wave functions. This effect enables Fermi transitions to nonanalog $0^+$ states, which are isospin forbidden. The spread of Fermi strength can be exactly calculated in valence spaces through diagonalization of the full Hamiltonian matrix. At the weak isospin-mixing limit, when it is sufficient to include only the isobaric analog state (typically the ground state) and the next $0^+$ state (denoted as $0^+_2$), $\delta_C$ is given by, 
\begin{equation}\label{2lev}
\delta_C\approx\frac{|(0^+_2||V_{INC}||f)|^2}{\Delta E^2}, 
\end{equation}
where $\Delta E$ represents the energy separation. 
Throughout this paper, we use round bracket for many-particle states with definite isospin. Although Eq.\,\eqref{2lev} is not always quantitatively justified, the strong dependence on $\Delta E$ observed in exact many-body calculations is well supported, even when $\delta_C$ is as large as 1\,\%\,\cite{Xthesis}. The double bars in Eq.\,\eqref{2lev} indicate the reduction in angular momentum. 

Since isospin mixing is principally driven by the infinite-range Coulomb interaction, it can extend far beyond the shell-model valence space, which is typically limited to a single oscillator shell. As a result, the correction values obtained from shell-model calculations are generally underestimated, except when the chosen effective Hamiltonian incorrectly yields a $\Delta E$ value that is too small. Ideally, the missing contribution can be partially captured through the construction of an effective transition operator. However, this first-principle approach requires the interaction matrix elements to be treated microscopically\footnote{It may also require a nonorthogonal isospin-dependent basis, since the isospin-raising operator is self-effective under isospin SU(2) symmetry.}. Within the phenomenological framework, we directly replace the isospin-invariant harmonic oscillator basis with realistic radial wave functions in the evaluation of the Fermi matrix element. Consequently, $\delta_C$ can be decomposed into two parts, 
\begin{equation}\label{c}
\delta_C \approx \delta_{C1} + \delta_{C2}, 
\end{equation} 
where $\delta_{C1}$ accounts for in-valence-space mixing induced by $V_{INC}$, and the ad-hoc term $\delta_{C2}$ arises from the fact that protons are less bound than neutrons, so the initial proton wave function imperfectly overlaps the final neutron one. 
Note that several higher-order terms, excluded from Eq.\,\eqref{c}, are insignificant, as demonstrated in Ref.\,\cite{PhysRevC.109.014317}. The effect of radial excitation, recently investigated in Ref.\,\cite{xayavong2025}, is also disregarded in the present work. The first leading-order term, $\delta_{C1}$, is calculated via, 
\begin{equation}\label{im}
\delta_{C1} = 2-\sqrt{2}\sum_{k_\alpha}\sqrt{j_\alpha+1} \braket{f||[a_{k_\alpha,n}^\dagger\otimes \tilde{a}_{k_\alpha,p}]^{(0)}||i} , 
\end{equation}
where $k$ stands for the spherical quantum numbers $nlj$. 
The one-body transition densities (the reduced matrix elements in Eq.\,\eqref{im}) must be evaluated with an accurate isospin nonconserving effective interaction. 
The other leading-order term, $\delta_{C2}$, can be arranged as
\begin{equation}\label{ro}
\begin{array}{ll}
\displaystyle \delta_{C2} & \displaystyle= \frac{1}{\sqrt{3}}\sum_{k_\alpha}\left[\sum_{\pi'}|(f|||a_{k_\alpha}^\dagger|||\pi')|^2 \Lambda_{k_\alpha}^{\pi'} \right. \\[0.1in]
&\displaystyle\left.- \frac{1}{2}\sum_{\pi''}|(f|||a_{k_\alpha}^\dagger|||\pi'')|^2 \Lambda_{k_\alpha}^{\pi''} \right] ,
\end{array}
\end{equation}
where $\pi'$ ($\pi''$) denotes intermediate $(A-1)$-nucleon states with lesser (greater) isospin, $T_{\pi'}=1/2$ ($T_{\pi'}=3/2$). The triple bars in the matrix elements indicate reductions in both angular momentum and isospin. The radial mismatch factor is given by 
\begin{equation}
\Lambda_{k_\alpha}^{\pi} = 1-\int_0^\infty R^\pi_{k_\alpha,p}(r)R^\pi_{k_\alpha,n}(r)r^2 dr, 
\end{equation}
with radial wave functions reproducing separation energies with respect to intermediate states. Note that $\Lambda_{k_\alpha}^{\pi}$ is always finite when evaluated using eigenfunctions of a realistic potential, where the decaying protons are slightly less bound than the corresponding neutrons. Additionally, the magnitude of $\Lambda_{k_\alpha}^{\pi}$ decreases as the excitation energy of the intermediate state increases. 

\begin{figure}
\centering
\includegraphics[]{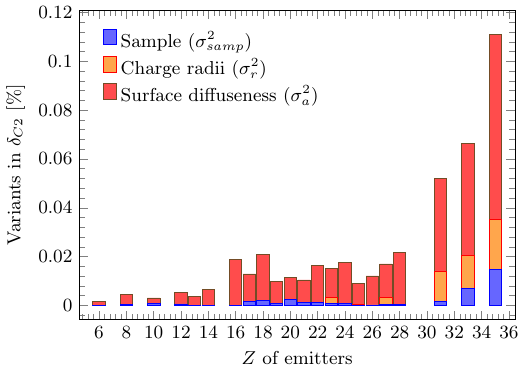}
    \caption{Uncertainty budgets for $\delta_{C2}$. See the main text for description. }
    \label{fig10}
\end{figure}

Generally, the cancellation between the lesser- and greater-isospin components in Eq.\,\eqref{ro} is incomplete, even for highly occupied orbitals, where the French-MacFarlane sum rule\,\cite{FrMac1961} implies, 
\begin{equation}
\sum_{\pi'}|(f|||a_{k_\alpha}^\dagger|||\pi')|^2 \approx \frac{1}{2}\sum_{\pi''}|(f|||a_{k_\alpha}^\dagger|||\pi'')|^2,  
\end{equation}
as $\Lambda_\alpha^{\pi'}$ is always larger than $\Lambda_\alpha^{\pi''}$, given that $\ket{\pi''}$ lies above $\ket{\pi'}$. Desirably, spectroscopic factors (SFs) deduced from transfer reaction cross sections can be directly substituted into Eq.\,\eqref{ro}. 
Unfortunately, experimental data are often available only for a few low-lying states without sufficient precision. Despite this limitation, the existing data can still be used as a guide for model space selection. For instance, if unnatural parity states appear at low energy in the spectrum of the $(A-1)$-nucleon spectator and carry substantial SFs, the associated core orbitals must be explicitly included. This approach was first suggested by Towner and Hardy (TH)\,\cite{ToHa2008} and has proven to be very useful--for example, in the case of $^{46}$V, where the $sd$-shell orbitals significantly contribute. In addition to this experimental guidance, our choice of model spaces for the present calculations also relies on the feasibility of full-space diagonalization. See Table\,\ref{tab1} for details on the inputs to the shell-model calculations. 

\begin{figure*}[ht!]
\centering
\includegraphics[]{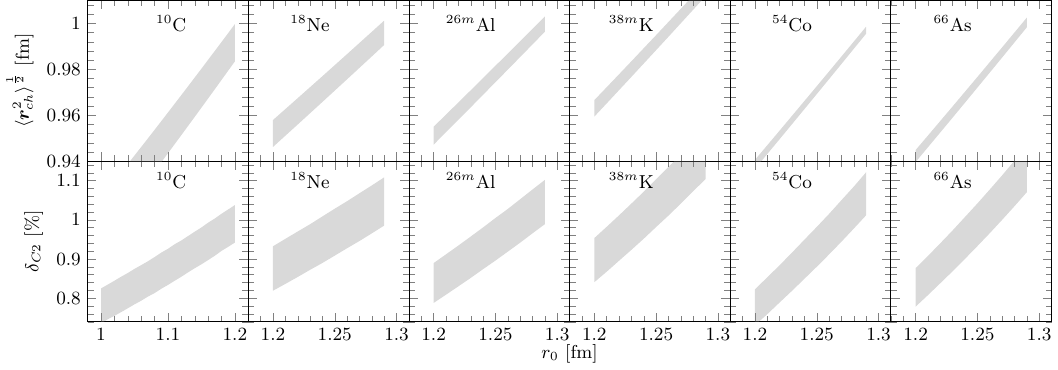}
\caption{Sensitivity of charge radii and $\delta_{C2}$ to the WS parameters. The charge radii are normalized to experimental values, and $\delta_{C2}$ is normalized to the values used in Ref.\,\cite{HaTo2020}. The bandwidths reflects the variation of the surface diffuseness parameter by $\pm0.1$\,fm around its standard value of 0.662~fm. }
\label{fig}
\end{figure*}

\section{$\delta_{C1}$ correction}\label{C1}

We explore $\delta_{C1}$ using $V_{INC}$ produced by OB\,\cite{OrBr1989x} for the $p$, ZBM, $sd$, and $pf$ spaces. For the ZBM2 and $jj44$ spaces, where no well-established $V_{INC}$ is available, we simply incorporate experimental IVSPEs and two-body Coulomb matrix elements. We notice that $V_{INC}$ interactions used by TH were evaluated on a case-by-case basis\,\cite{ToHa2008,ToHa2002}. In addition to the Coulomb and IVSPEs, they also modified the $T=1$ proton-neutron matrix elements, constrained by experimental IMME coefficients. 
Our calculations are performed using the code NuShellX@MSU~\cite{NuShellX} and the computational facility at Mésocentre de Calcul Intensif Aquitain (MCIA). In several cases, TH's model spaces cover more orbitals than ours, as summarized in Table\,\ref{tab1}, but typically involve a severe truncation (this is mentioned in Ref.~\cite{ToHa2008}, without providing further details). For instance, they suggested including the $d_{5/2}$ orbital for $A=42$ and $46$ to account for the presence of $5/2^+$ states in the spectra of the corresponding intermediate systems. We adopt the ZBM2 space for this mass region, where a well-established $H_0$ is available~\cite{Bis2014}, and full-space diagonalization has become manageable on supercomputers. In our opinion, this choice is reasonably justified, as the $5/2^+$ states typically lie above 5\,MeV, leading to a smaller radial mismatch factor. Moreover, the experimentally deduced SFs associated with these unnatural parity states are not particularly large. It should also be noted that introducing truncated $d_{5/2}$ would induce larger center of mass contamination and, consequently, additional uncertainties in the calculations. 
In general, $\delta_{C1}$ is strongly interaction-dependent. As an empirical refinement, calculated $\delta_{C1}$ values may be scaled with experimental $\Delta E^2$, in accordance with Eq.\,\eqref{2lev}. Although the relation is derived at the weak isospin-mixing limit, this scaling is usually acceptable, given that the magnitude of $\delta_{C1}$ is typically below 0.2\,\% (see, for example, Ref.\,\cite{HaTo2020}). In this study, we follow the same procedure and consider the average between unscaled and scaled $\delta_{C1}$ values. 
Our results for $\delta_{C1}$ are given in Table\,\ref{tab2}. The reported uncertainties stem from the use of different Hamiltonians for the same nucleus within the same model space, as well as from the difference between scaled and unscaled values. One can see that our results are in remarkable agreement with those of TH~\cite{HaTo2020,HaTo2015x,ToHa2008} in most cases. An exception is our $\delta_{C1}$ value for $^{10}$C, which is an order of magnitude larger
, despite both being obtained in the $p$ shell. This discrepancy comes partly from a larger mixing matrix element and partly from a lower position of the $0_2^+$ state in $^{10}$B, consistent with Eq.\,\eqref{2lev}. For a similar reason, our $\delta_{C1}$ value for $^{22}$Mg is also slightly larger. Conversely, we obtain significantly smaller $\delta_{C1}$ values for $^{30}$S, $^{34}$Cl, $^{34}$Ar and $^{62}$Ga, suggesting another issue opposite to that faced in $^{10}$C. 
Overall, as seen from Table~\ref{tab2}, $\delta_{C1}$ is very sensitive to the details of the shell-model interaction (even after energy scaling) and, in particular, to the charge-dependent $V_{INC}$ term. Nevertheless, its contribution is no more than 10~\% to the total correction. 

\begin{table*}[ht!]
\label{tab2}
\caption{List of the calculated $\delta_{C1}$, $\delta_{C2}$ and the nucleus-independent $\mathcal{F}t$ values. Data on charge radii, and proton and neutron separation energies~\cite{AME2012} used to constrain WS radial wave functions are given in column 2, 3 and 4, respectively. The experimental $ft$ values are taken from Ref.\,\cite{PhysRevC.109.045501} when available, or from Ref.~\cite{HaTo2020} otherwise. The radiative correction values are the same as those used in Ref.~\cite{HaTo2020}.}
\begin{threeparttable}
\begin{ruledtabular}
\begin{tabular}{cccc|ccc|ccc}
\multirow{2}{*}{Emitters} & \multirow{2}{*}{$\braket{\bm{r}^2_{ch}}^\frac{1}{2}$~[fm]} & \multirow{2}{*}{$S_p$~[MeV]}  & \multirow{2}{*}{$S_n$~[MeV]} & \multicolumn{3}{c|}{This work} &  \multicolumn{3}{c}{TH~\cite{HaTo2020}}  \\
\cline{5-10}
	& &   &  & $\delta_{C1}~[\%]$ & $\delta_{C2}~[\%]$ & $\mathcal{F}t~[s]$ & $\delta_{C1}~[\%]$ & $\delta_{C2}~[\%]$ & $\mathcal{F}t~[s]$    \\
\hline
$^{10}$C\tnote{a} & 
2.638(36)\,\cite{ohayon2025}  & 4.007  & 8.437 & 0.102(45) & 0.227(42) & 3073.8(45) & 0.010(10) & 0.165(15)& 3075.7(44) \\
$^{14}$O\tnote{a} & 
2.705(10)\,\cite{ohayon2025}  & 4.627 & 10.553 & 0.039(21) & 0.303(66) & 3069.3(28) & 0.055(20) & 0.275(15)& 3070.2(19) \\
$^{18}$Ne& 2.934(10)\,\cite{ohayon2025} 
& 3.923 & 9.150 &  0.086(30)  & 0.217(55) & 2935.2(797) & 0.155(30)  & 0.405(25) & 2930(80) \\
$^{22}$Mg\tnote{a} & 3.0691(7)\cite{PhysRevLett.108.042504}  & 5.504 & 11.068 & 0.032(1) & 0.364(74) & 3075.8(74) & 0.010(10) & 0.370(20)& 3076.2(70) \\
$^{26}$Si\tnote{a} & 
3.136(04)\,\cite{ohayon2025}  & 5.514 & 11.365 & 0.025(4) & 0.384(80) & 3074.5(62) & 0.155(20) & 0.405(25)& 3075.4(57) \\
$^{30}$S& 
3.226(06)\,\cite{ohayon2025}  & 4.396 & 11.319 & 0.059(4) & 0.661(137) & 3027.0(414) & 0.155(20) & 0.700(20)& 3027(41) \\
$^{34}$Ar\tnote{a} & 
3.365(11)\,\cite{ohayon2025} & 4.664 & 11.508 & 0.010(1) & 0.648(145) & 3073.6(53) & 0.030(10) & 0.665(55)& 3075.1(31) \\
$^{38}$Ca\tnote{a} & 3.4652(19)\,\cite{PhysRevResearch.2.022035}  & 4.547 & 12.072 & 0.012(1) & 0.604(106) & 3080.9(69) & 0.020(10) & 0.745(70)& 3077.8(62) \\
$^{42}$Ti& 
3.576(05)\,\cite{ohayon2025}  & 3.751 & 11.550 &   & 0.614(127) & 3099.1(885) & 0.105(20) & 0.835(75)& 3097(88) \\
$^{46}$Cr& 
3.670(05)\,\cite{ohayon2025}  & 4.882 & 13.265 & 0.025(8) & 0.656(132) & 3143.2(1006) & 0.045(20) & 0.715(85)& 3141(100)\\
$^{50}$Fe& 
3.709(04)\,\cite{ohayon2025} & 4.141 & 13.062 & 0.032(3) & 0.537(109) & 3116.1(715) & 0.025(20) & 0.635(45) & 3118(72) \\
$^{54}$Ni& 
3.737(3)\,\cite{PhysRevLett.127.182503} & 3.853 & 13.422 & 0.029(4) & 0.666(111) & 3079.3(505) & 0.065(30) & 0.725(60) & 3077(50) \\ 
\hline
$^{26m}$Al\tnote{a} & 3.130(15)\,\cite{PhysRevLett.131.222502}  & 6.306 & 11.093 & 0.007(1) & 0.279(59) & 3071.3(21) & 0.030(10) & 0.280(15)& 3072.4(11) \\
$^{34}$Cl\tnote{a} & 
3.327(4)\,\cite{ohayon2025} & 5.143 & 11.417 & 0.052(4) & 0.525(112) & 3069.8(36) & 0.100(10) & 0.550(45)& 3071.6(18) \\
$^{38m}$K\tnote{a} & 
3.4353(29)\,\cite{PhysRevA.111.042813} & 5.142 & 11.839 & 0.094(3) & 0.424(99) & 3075.0(33) & 0.105(20) & 0.565(50)& 3072.9(20) \\ 
$^{42}$Sc\tnote{a} & 
3.558(30)\,\cite{PhysRevLett.134.182501} & 11.481 & 1.451& 0.035(20)& 0.480(103) & 3074.8(35) & 0.020(10) & 0.645(55)& 3071.7(20) \\
$^{46}$V\tnote{a} &  
\emph{3.642(5)}\,\cite{ohayon2025} & 5.354 & 13.189 & 0.072(24) & 0.541(113) & 3074.4(37) & 0.075(30) & 0.545(55)& 3074.3(20) \\
$^{50}$Mn\tnote{a} &  3.728(41)\,\cite{ohayon2025} & 4.584 & 13 & 0.041(2) & 0.466(97) & 3073.7(33) & 0.035(20) & 0.630(25)& 3071.1(16) \\ 
$^{54}$Co\tnote{a} & 
\emph{3.714(4)}\,\cite{ohayon2025} & 4.351 & 13.378 & 0.021(1) & 0.523(119) & 3077.1(41) & 0.050(30) & 0.720(60)& 3070.4(25)\\
$^{62}$Ga\tnote{a} &  
\emph{3.906(6)}\,\cite{ohayon2025} & 2.927 & 12.890 & 0.05(3) & 0.927(198) & 3081.0(66) & 0.275(55) & 1.200(200)& 3072.4(67) \\
$^{66}$As& 4.02(10)\,\cite{ToHa2002}  & 2.836 & 13.2 & 1.087(195)& 1.048(258) & &   & 1.350(400)&   	\\
$^{70}$Br& 4.10(10)\,\cite{ToHa2002}  & 2.280 & 13.566 & 1.559(258)& 1.391(333) & &   & 1.250(250)&   	\\
\hline
\multicolumn{4}{r|}{Average (best 14), $\overline{\mathcal{F}t}$} & 	\multicolumn{3}{c|}
{$3073.11(99)_{stat}(36)_{\delta_R'}(173)_{\delta_{NS}}$}	& \multicolumn{3}{c}{$3072.23(57)_{stat}(36)_{\delta_R'}(173)_{\delta_{NS}}$} \\
\multicolumn{4}{r|}{$\chi^2/\nu$} & 	\multicolumn{3}{c|}{0.624} & \multicolumn{3}{c}{0.492}  \\
\multicolumn{4}{r|}{$|V_{ud}|$}   & 	\multicolumn{3}{c|}{0.97359(33)} & \multicolumn{3}{c}{0.97372(31)}  \\
\end{tabular}
\end{ruledtabular}
\begin{tablenotes}
{ \raggedright
\item[a] Transitions used to obtain $\overline{\mathcal{F}t}$. \\
}
\end{tablenotes}
\end{threeparttable}
\end{table*}

\section{$\delta_{C2}$ correction}\label{C2}

The Hamiltonians $H_0$ used in our calculations of $\delta_{C2}$ are listed in the fourth column of Table\,\ref{tab1}. 
To evaluate $\Lambda_{k_\alpha}^{\pi}$, we consider two different parameter sets of the WS potential: one being close to that of Ref.~\cite{BohrMott} (denoted as BM$_m$), while the other, referred to as SWV, being based on Ref.~\cite{SWV}. The details of these parameterizations and the role of each term are discussed extensively in Refs.~\cite{XaNa2018,Xthesis}. Briefly speaking, we use the potential of the form, 
\begin{equation}\label{WS}
\displaystyle V(r) = V_{0}f_0(r) + V_{s}\left(\frac{r_s}{\hbar}\right)^2 \frac{1}{r}\frac{d}{dr}f_s(r)
\braket{\boldsymbol{l}\cdot \boldsymbol{\sigma}} + V_{c}(r),
\end{equation}
where $\displaystyle f_i(r)=\{1+\exp{[{(r-R_i)}/{a_i}]}\}^{-1}$ with $i$ denoting either $0$ for the central term or $s$ for the spin-orbit term. The radius is modeled as $R_i = r_i\times(A-1)^{1/3}$. 
The Coulomb term is determined under the assumption of a uniformly charged sphere, with its radius constrained by the measured charge radius of the parent nucleus, as described in Refs.~\cite{Elton,XaNa2018}. Since our previous work on $sd$-shell nuclei~\cite{XaNa2018}, our treatment of charge radii has extended beyond the closure approximation. Specifically, the square of the charge radius is given by the expectation value of $\bm{r}^2$ operator in the ground state of the parent nucleus, where the occupation numbers are replaced by the sum of SFs obtained from shell model calculations. Essentially, a dependence of radial wave functions on intermediate states is introduced, similar to Eq.\,\eqref{ro}. The generalized formula reads, 
\begin{eqnarray}\label{radius}
\braket{\bm{r}^2}=\frac{1}{Z} \sum_{k_\alpha\pi} |\braket{i||a^{\dagger}_{k_\alpha,p} ||\pi}|^2 
\braket{k_{\alpha,p}|| \bm{r}^2 ||k_{\alpha,p}}^\pi, 
\end{eqnarray}
where 
the proton occupancy of the shell-model core orbitals is taken as $(2j+1)$. In practice, various corrective terms~\cite{XaNa2018}, particularly those accounting for finite size and spurious center-of-mass contributions, must be applied to Eq.~\eqref{radius}. The charge-radius data used in the present calculations are listed in the second column of Table\,\ref{tab2}. We emphasize that Eq.\,\eqref{radius} not only accounts for many-body correlations through shell-model calculations of SFs, but also enables a self-consistent adjustment of two parameters, since the radial wave functions in Eq.\,\eqref{radius} are naturally the same as those used in Eq.\,\eqref{ro}. Note that TH used a traditional approach for evaluating the $\bm{r}^2$ operator. The effect of this charge-radius generalization is not uniform; in the most significant case, it reduces $r_0$ in $^{18}$Ne from 1.29~fm to 1.25~fm, which subsequently leads to a reduction in $\delta_{C2}$ by approximately 0.18~\%.  
Our adjustment procedure can be summarized as follows. The spin-orbit term parameters and $a_0$ are always fixed at their standard values. Two approaches are considered. In the first, we vary $V_0$ and $r_0$ to reproduce one-nucleon separation energies (for protons and neutrons separately) and charge radii, respectively. In the second, we fix $V_0$ at the value obtained from the first approach for the ground state and introduce a separate surface-peaked term, 
\begin{equation}\label{g}
V_g(r)=\left(\frac{\hbar}{m_{\pi}c}\right)^2\frac{V_g}{a_sr}\exp\Big(\frac{r-R_s}{a_s}\Big) [ f_s(r) ]^2 
\end{equation}
for excited states, where $V_g$ is a free parameter. The parameters $R_s$ and $a_s$ are same as in Eq.\,\eqref{WS}. As we have found out for the $sd$-shell cases~\cite{XaNa2018}, the procedure involving the variation of $V_0$ essentially removes the dependence on a particular parameterization: the BM$_m$ and SWV results are very similar. Contrary, the fit of the surface term results in slightly larger differences between the two parameterizations, and we keep this in our analysis. The other surface-peaked term considered by TH~\cite{ToHa2008} is disregarded, as in certain cases it introduces an undesired correlation (often quadratic) between the charge radii and $r_0$.  
We find that the correlation of $\delta_{C2}$ with $r_0$ and $a_0$ can be approximated by $\delta_{C2}(i)\approx C(i) + S_r (i) r_0(i) + S_a(i) a_0(i)$ (see Fig.\,\ref{fig}), where $i$ labels an individual calculation defined by a specific choice of effective interaction, WS parameter set, and the inclusion of the surface-peaked term. 
Accordingly, the associated uncertainty can be evaluated as $\sigma_{WS}^2(i)\approx S_r^2 (i)\sigma_r^2(i) + S_a^2(i) \sigma_a^2(i)$, where the slopes $S_r(i)$ and $S_a(i)$ are numerically obtained for each transition. The optimal $r_0$ value and its uncertainty $\sigma_r(i)$ are extracted from the charge-radius data, similar to approach described in Ref.\,\cite{XaNa2018}. Since there is no constraint for $a_0$, we assign $\sigma_a(i)=0.1$\,fm to reflect its typical range found in the literature\,\cite{SWV}. This procedure generates the data set $\{\delta_{C2}(i)\pm\sigma_{WS}(i)\}$. Since the spread within the sample is much smaller than $\sigma_{WS}(i)$, a weighted average would lead to a drastically small uncertainty. To avoid this reduction, we adopt the prescription $\sigma_{C2}^2=\sigma_{samp}^2+\bar{\sigma}_{WS}^2$, where $\bar{\sigma}_{WS}$ denotes the average of $\sigma_{WS}(i)$ and $\sigma_{samp}$ is the sample standard deviation of the central values $\delta_{C2}(i)$. Our uncertainty budget to $\delta_{C2}$ is depicted in Fig.\,\ref{fig10}. 
For the present calculations, at least 100 intermediate states are taken into account for each spin-parity in all transitions. 
The convergence of charge radii within Eq.\,\eqref{radius} is generally faster. 

Our results for $\delta_{C2}$ are listed in the sixth column of Table~\ref{tab2}. 
For comparison, the $\delta_{C2}$ values taken from the latest TH survey~\cite{HaTo2020} are shown in the ninth column. We observe that, within a given model space, the spread arising from the use of different effective interactions is typically marginal, even when some chosen interactions do not reproduce the energy spectra or electromagnetic properties with compatible quality. 
For $A\le34$, our $\delta_{C2}$ values are in good agreement with those of TH, despite the use of different model spaces and effective interactions in the region around $^{16}$O. A clear exception is $^{18}$Ne, for which our value is smaller by approximately a factor of two, due to the updated data and generalized expression for charge radii. The sensitivity of $\delta_{C2}$ to $r_0$ and $a_0$ is depicted in Fig.\,\ref{fig}. For heavier nuclei, apart from the emitters with $A=46$ and $70$, our results tend to exhibit a systematically reduction compared to those of TH, although both remain consistent within the error bars in most cases. This discrepancy is primarily caused by the differences in the adopted model spaces and effective interactions. 

\section{Standard model implications}\label{impli}

The corrected $\mathcal{F}t$ values are evaluated for the 14 best-known transitions, as well as for other potential candidates from $^{10}$C to $^{70}$Br. 
We exclude $^{74}$Rb, for which an untruncated calculation within the $jj44$ space remains challenging due to the large number of required intermediate states. However, since calculations for a limited number of low-lying states have been performed in this region~\cite{PhysRevC.92.024320}, a good approximation to the present problem might be obtained using the Lanczos strength-function method. This case will be addressed in a separate study.  
Since our calculations of the isospin mixing term do not reproduce the relevant experimental data well in several cases, we adopt the $\delta_{C1}$ values of TH for our analysis, as they provide remarkable agreement with data from isospin-forbidden transitions (see Ref.~\cite{ToHa2008} and Refs. therein). Data on the radiative corrections, $\delta_{NS}$, $\delta_R'$, and $\Delta_R^V$, are identical to those used in Ref.~\cite{HaTo2020}. The experimental $ft$ values are taken from Ref.~\cite{PhysRevC.109.045501} when available, or from Ref.\,\cite{HaTo2020} otherwise. The results for the individual transitions are listed in the eighth column of Table~\ref{tab2}. The averaged $\overline{\mathcal{F}t}$ and its $\chi^2/\nu$ (measure of local fluctuation) are shown in the last row of the same Table. Here we adopt the evaluation procedure from Ref.\,\cite{HaTo2020}, so that the uncertainty on $\delta_R'$ is not incorporated within the individual $\mathcal{F}t$ values but, it is directly added to the averaged $\overline{\mathcal{F}t}$ as a systematic uncertainty. The other source of the systematic uncertainty originates from $\delta_{NS}$ as described in detail in Ref.\,\cite{HaTo2020}. From our $\delta_{C2}$ correction set, we obtained $\overline{\mathcal{F}t}=3073.11(99)_{stat}(36)_{\delta_R'}(173)_{\delta_{NS}}$ and $|V_{ud}|=0.97359(33)$. It is interesting to mention that our central $\overline{\mathcal{F}t}$ value is $0.32\sigma$ larger and our $|V_{ud}|$ value is $0.29\sigma$ smaller, in comparing with those extracted from the latest calculations of TH~\cite{HaTo2020}. Additionally, our result yields a $\chi^2/\nu$ value of 0.624 which is slightly larger than 0.492 the value obtained by TH. This suggests that the corrected $\mathcal{F}t$ values produced in the present work are less consistent with the Standard Model although the additional uncertainties arising from the surface diffuseness parameter are incorporated for $\delta_{C2}$. This discrepancy is seemingly attributed to the lowering of our $\delta_{C2}$ values for $A\ge38$ as a consequence of the different configuration spaces. 

\begin{figure}[ht!]
\centering
\includegraphics[]{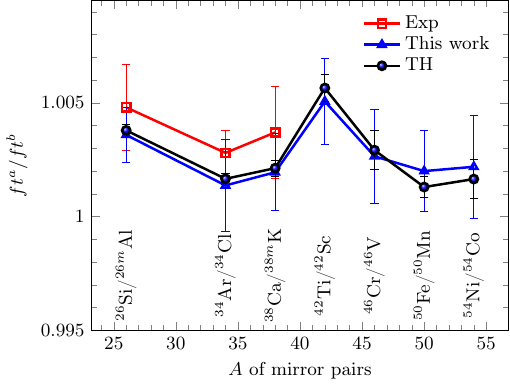}
\caption{Comparison of the mirror $ft$ ratio for $A = 26, 34, 38, 42, 50$, and $54$. The Exp and TH values are taken from Ref.\,\cite{HaTo2020}.}
\label{fig3}
\end{figure}

Assuming valid CVC, the ratio of $ft$ values for a pair of mirror superallowed transitions can be written as 
\begin{equation}\label{mirror}
\frac{ft^a}{ft^b}=1+(\delta_R^{'b}-\delta_R^{'a})+(\delta_{NS}^b-\delta_{NS}^a)-(\delta_{C}^b-\delta_{C}^a), 
\end{equation}
where superscript $a$ and $b$ denote the decay of $T_z=-1$ and $T_z=0$ nuclei, respectively. The advantage offered by Eq.\,\eqref{mirror} is that the theoretical uncertainty on the difference $(\delta_{C}^b-\delta_{C}^a)$ is significantly reduced compared to the uncertainties on $\delta_{C}^b$ and $\delta_{C}^a$ individually. Interestingly, the ratio of mirror $ft$ values is very sensitive to $\delta_{C}$, and hence can be used to test the merits of the available calculations. In particular, the difference in this correction between emitters with $T_z=-1$ and $T_z=0$, arises mainly from the radial mismatch contribution, due to the binding energy effect, as emphasized in Ref.\,\cite{PhysRevC.105.044308}. We have extracted the $ft$ ratio for 7 mirror transitions from $A=26$ to 54, the results are displayed in Fig.~\ref{fig3} in comparison with those of TH~\cite{HaTo2020} and the available experimental values. It is seen that the present results are in fairly good agreement with the other two, although our $\delta_{C2}$ values for $A\ge38$ tend to be smaller than those of TH. 

\section{Conclusion}\label{con} 

We carry out shell-model calculations of the isospin-symmetry breaking correction to superallowed $0^+\rightarrow0^+$ nuclear $\beta$ decay of isotriplets, with special emphasis on the dominant radial mismatch term $\delta_{C2}$. 
All potential candidate nuclei are included, except for $^{74}$Rb, where the diagonalization of the Hamiltonian matrix in the full $jj44$ model space is still not feasible, even at high-performance computing facilities. Our evaluation of the isospin mixing term $\delta_{C1}$ is based on existing charge-dependent effective interactions produced via a global fit or if unavailable, we simply add two-body Coulomb interaction and isovector single-particle energies on top of a well-established effective isoscalar interactions. The obtained $\delta_{C1}$ values agree with the previous study for some cases, while discrepancies appear in the other cases, typically when the low-lying $0^+$ levels are not quantitatively reproduced. 
In addition to the use of untruncated model spaces, several improvements are introduced for the calculations of $\delta_{C2}$. These comprise the use of updated data and generalized formula for charge radii, which serve as a constraint for radial wave functions. This generalization enables a self-consistent fitting for two parameters, such as depth and radius. While we do not have a robust constraint for the surface diffuseness parameter, a fluctuation of $\pm0.1$\,fm around a central value is incorporated and now represents the dominated uncertainty source for $\delta_{C2}$. 
Our $\delta_{C2}$ values agree well with those of Towner and Hardy for light nuclei below $A=34$, with the exception of $^{18}$Ne. For heavier nuclei, however, our results tend to underestimate the Towner-Hardy values, which is likely due to differences in configuration spaces and effective interactions employed. The averaged $\mathcal{F}t$ value obtained in the present study shows slightly reduced consistency with the CVC, as indicated by a somewhat larger normalized $\chi^2$. The extracted value of $|V_{ud}|$ slightly increases the deviation from unity in the CKM sum rule, thought it remains consistent with the Towner-Hardy results within our newly assigned uncertainty. 

To continue, it is desirable to further enlarge the model space within the shell-model framework and to develop accurate effective interactions using state-of-the-art computing technologies. 
The reliability of the present approach seems to be questionable for light nuclei, where the WS eigenfunctions become increasingly sensitive to subtle details of the potential, especially the Coulomb term. Moreover, the spurious center-of-mass contamination, which cannot be entirely suppressed when using a realistic potential, tends to be more severe in light system. A rigorous investigation of these aspects would be valuable. 

\begin{acknowledgments} 
Discussions with M.~Bender, B.~Blank and T.~Kurtukian-Nieto are gratefully acknowledged. L.~Xayavong would like to thank CENBG for its hospitality during his visit in 2019. The work was supported by IN2P3/CNRS, France. 
L. Xayavong is also supported the National
Research Foundation of Korea(NRF) grant funded by the
Korea government(MSIT)(No. RS-2024-00457037) and by Global-Learning \& Academic research institution for Master’s Ph.D. students, and Postdocs (LAMP) Program of the National Research Foundation of Korea (NRF) grant funded by the Ministry of Education (No. RS-2024-00442483).  
\end{acknowledgments}

\bibliography{prl}

\end{document}